\documentclass[%
 aip,
 amsmath,amssymb,
 reprint,%
]{revtex4-1}

\usepackage{graphicx}
\usepackage{subcaption} 
\usepackage{dcolumn}
\usepackage{bm}
\usepackage{ulem} 

\usepackage[utf8]{inputenc}
\usepackage[T1]{fontenc}
\usepackage{mathptmx}
\usepackage{etoolbox}

\usepackage{chemformula} 
\usepackage{multirow} 

\makeatletter
\def\@email#1#2{%
 \endgroup
 \patchcmd{\titleblock@produce}
  {\frontmatter@RRAPformat}
  {\frontmatter@RRAPformat{\produce@RRAP{*#1\href{mailto:#2}{#2}}}\frontmatter@RRAPformat}
  {}{}
}%
\makeatother

\begin{document}

\preprint{AIP/123-QED}

\title[The Role of the Dopant in the Electronic Structure of Erbium-Doped \ch{TiO2} for Quantum Emitters]{The Role of the Dopant in the Electronic Structure of Erbium-Doped \ch{TiO2} for Quantum Emitters}
\author{J. B. Martins}
    \email{jbarbosamartins@anl.gov}
\affiliation{ 
X-ray Science Division, Argonne National Laboratory, Lemont, Illinois 60439, USA
}%
\author{G. Grant}%
 \altaffiliation[Also at ]{Materials Science Division, Argonne National Laboratory, Lemont, Illinois 60439, USA}
\affiliation{ 
Pritzker School of Molecular Engineering, University of Chicago, Chicago, Illinois 60637, USA
}%

\author{D. Haskel}
\affiliation{ 
X-ray Science Division, Argonne National Laboratory, Lemont, Illinois 60439, USA
}%

\author{G. E. Sterbinsky}
\affiliation{ 
X-ray Science Division, Argonne National Laboratory, Lemont, Illinois 60439, USA
}%

\author{I. Masiulionis}%
 \altaffiliation[Also at ]{Materials Science Division, Argonne National Laboratory, Lemont, Illinois 60439, USA}
\affiliation{ 
Pritzker School of Molecular Engineering, University of Chicago, Chicago, Illinois 60637, USA
}%

\author{K. Sautter}%
 \altaffiliation[Also at ]{Materials Science Division, Argonne National Laboratory, Lemont, Illinois 60439, USA}
\affiliation{ 
Pritzker School of Molecular Engineering, University of Chicago, Chicago, Illinois 60637, USA
}%

\author{E. Karapetrova}
\affiliation{ 
X-ray Science Division, Argonne National Laboratory, Lemont, Illinois 60439, USA
}%

\author{S. Guha}
 \altaffiliation[Also at ]{Materials Science Division, Argonne National Laboratory, Lemont, Illinois 60439, USA}
\affiliation{ 
Pritzker School of Molecular Engineering, University of Chicago, Chicago, Illinois 60637, USA
}%

\author{J. W. Freeland}
\affiliation{ 
X-ray Science Division, Argonne National Laboratory, Lemont, Illinois 60439, USA
}%

\date{\today}

\begin{abstract}
Erbium-doped \ch{TiO2} materials are promising candidates for advancing quantum technologies, necessitating a thorough understanding of their electronic and crystal structures to tailor their properties and enhance coherence times. This study explored epitaxial erbium-doped rutile \ch{TiO2} films deposited on r-sapphire substrates using molecular beam epitaxy. Photoluminescence excitation spectroscopy demonstrated decreasing fluorescence lifetimes with erbium doping, indicating limited coherence times. Lattice distortions associated with \ch{Er^{3+}} were probed by X-ray absorption spectroscopy, indicating that erbium primarily occupies \ch{Ti^4+} sites and influences oxygen vacancies. Significant lattice distortions in the higher-order shells and full coordination around erbium suggest that  additional defects are likely prevalent in these regions. These findings indicate that defects contribute to limited coherence times by introducing alternative decay pathways, leading to shorter fluorescence lifetimes.

\end{abstract}

\maketitle

\section{\label{sec:level1}Introduction}

The relentless growth in data storage and the continuous trend of device miniaturization has driven technological progress to approach the quantum limit. \cite{Awschalom:2018p516, Laucht:2021p162003} Quantum Information Science (QIS), grounded in the principles of quantum mechanics, emerges as a groundbreaking solution for efficient data handling, secure information exchange, and long-distance communication while ensuring data integrity. \cite{Acin:2018p80201, DiVincenzo:1995p255} At the heart of quantum technologies lies the manipulation of qubits, the fundamental units of quantum information.\cite{Zhang:2020p31308, DiVincenzo:1995p255} 
Quantum coherence plays a crucial role in realizing quantum technologies\cite{Wolfowicz:2021p906}, which enables qubits to maintain the superposition of quantum states. However, the challenge lies in preserving coherence which is limited by the inherent susceptibility to environmental influences. Therefore, ensuring the prolonged persistence of quantum states in superposition (i.e., long coherence time) is essential for successfully implementing quantum information technologies. \cite{Wolfowicz:2021p906,Zhang:2020p31308}

Solid-state spin qubits show significant promise among the candidates for QIS applications, owing to their compatibility with conventional microelectronics.  \cite{Wolfowicz:2021p906,Zhang:2020p31308} Nevertheless, the properties of solid-state spin qubits are linked to their crystal host, necessitating material engineering to enhance the quantum emitter's properties and mitigate decoherence sources like charge carriers or phonons. Among potential candidates, rare-earth ions (REIs) in solid-state systems stand out. \cite{Zhong:2019p2003} Specifically, the trivalent erbium ion (\ch{Er^{3+}}) is noteworthy for its optical 4f–4f transition within telecom-C band, aligning with the low-loss transmission in optical fibers, making it an ideal quantum emitter for quantum network systems. \cite{Boettger:2009p115104,Serrano:2022p241,Dibos:2022p6530,Rancic:2018p50,Ji:2023p,Stevenson:2022p224106} Selecting the optimal crystal host for REIs in QIS involves considering factors such as low nuclear and electron spin concentrations, nonpolar symmetry, and a large band gap. \cite{Wolfowicz:2021p906,Xie:2021p54111,Stevenson:2022p224106} These factors are instrumental in not only enhancing optical transitions but also in mitigating local effects in the vicinity of REIs, thereby addressing coherence issues effectively. Previous studies identified \ch{TiO2} as a promising host for \ch{Er^{3+}}, with the rutile and anatase phases being particularly stable. \cite{Phenicie:2019p8928, Shin:2022p81902} Recent research on Er-doped \ch{TiO2} films and bulk crystals has demonstrated narrow optical inhomogeneous linewidths, underscoring the importance of buffer and capping layers in enhancing optical properties. \cite{Phenicie:2019p8928, Singh:2022p}. Despite these advancements, further understanding is needed on how \ch{Er^{3+}} dopants influence the \ch{TiO2} host lattice's electronic and crystal structure.

This investigation focuses on epitaxial erbium-doped rutile \ch{TiO2} thin films grown on \textit{r}-plane sapphire substrates, aiming to explore how erbium incorporation affects the material's coherence times. We employed spectroscopy and diffraction techniques to correlate structural modifications with variations in coherence times. The selection of sapphire as the substrate was deliberate, as it enables the formation of high-quality single-crystal rutile \ch{TiO2}. X-ray diffraction (XRD) confirmed the persistence of single crystallinity despite doping. Variations on the \ch{Er^{3+}} fluorescence lifetimes, which set the upper bounds for coherence times, were assessed using photoluminescence excitation (PLE) spectroscopy revealing a downward trend with increasing doping. Additionally, information on local coordination, lattice distortion, and chemical environment near erbium sites was probed using X-ray absorption spectroscopy (XAS) and extended X-ray absorption fine structure (EXAFS). The analyses indicate that doping increases oxygen vacancy defect density within \ch{TiO2}. Furthermore, they suggest that erbium ions are fully coordinated, indicating redistribution of oxygen vacancies away from the \ch{Er} sites. This increase in additional defect density is likely to limit coherence by introducing alternative decay pathways upon \ch{Er^{3+}} excitation.

\section{Methods}

\subsection{Samples Growth}

Epitaxial Er:\ch{TiO2} thin films were deposited using a Riber C21 DZ Cluster molecular beam epitaxy (MBE) system. Titanium tetraisopropoxide (TTIP) from Sigma-Aldrich with a purity of 99.999\% (trace metal basis) was the titanium precursor as in the TTIP-based MBE of \ch{TiO2} detailed elsewhere. \cite{Jalan:2009p230} Growths were carried out on a r-plane (102) sapphire substrate at 640 ºC temperature, oxygen partial pressure of $7\times10^{-6}$ torr, and a TTIP beam-equivalent flux of $3.5\times10^{-6}$.  The growth rate was 32.5 nm per hour. \textit{In-situ} erbium (\ch{Er}) doping (20, 200, and 500 ppm) using metallic \ch{Er} was carried out by a high-temperature (860 ºC to 990 ºC) effusion cell.

\subsection{Structural and optical characterization}

Single crystal X-ray diffraction characterization was carried out at the 33BM beamline of the Advanced Photon Source (APS) at a photon energy of 20 keV. PLE spectroscopy was performed in a custom confocal microscopy setup designed for telecom C-band measurement of thin films, with samples mounted in a cryostat at 3.5 K (s50 Cryostation, Montana Instruments). A tunable $1.5$ $\mu$m laser illuminated the samples for 1.5 ms using pulses shaped by acousto-optic modulators. Emission from the samples subsequent to excitation was then collected by a Quantum Opus superconducting nanowire single photon detector (SNSPD) for 3 ms collection intervals. Sweeping the excitation laser wavelength resulted in PLE spectrum and excited state lifetime measurements at each wavelength for each measured sample. Additional information on this experimental setup is given in detail elsewhere \cite{Ji:2023p}. XAS measurements were conducted at \ch{O} \textit{K}-edge, \ch{Ti} \textit{L}-edge, and \ch{Er} \(M_5\)-edges at the 29ID beamline at the Advanced Photon Source (APS). The beam was directed along surface normal, and the spectra were averaged over the acquired horizontal and vertical polarization to ensure orientation-independent assessment of the local environment. The EXAFS measurements were performed at the 20-BM beamline of the APS. Data collection was conducted in a grazing incidence geometry, with an angle of incidence of a few degrees and with linear x-ray polarization in the plane of the film. The sample was rotated about the film surface normal to average in-plane polarizations. The temperature was maintained at 300 K throughout the entire experiment. Since the [101] direction is along the surface normal, separate FEFF8 simulations were carried out with linear polarization along orthogonal in-plane [-101] and [010] directions to obtain theoretical phase shifts and scattering amplitudes \cite{Ankudinov:1998p7565}.

\section{Results and Discussion}

\subsection{Characterization of epitaxial \ch{Er:TiO2} films by single crystal X-ray diffraction}

\begin{figure}[ht]
\begin{minipage}[b]{0.95\linewidth}
\centering
\includegraphics[width=\textwidth]{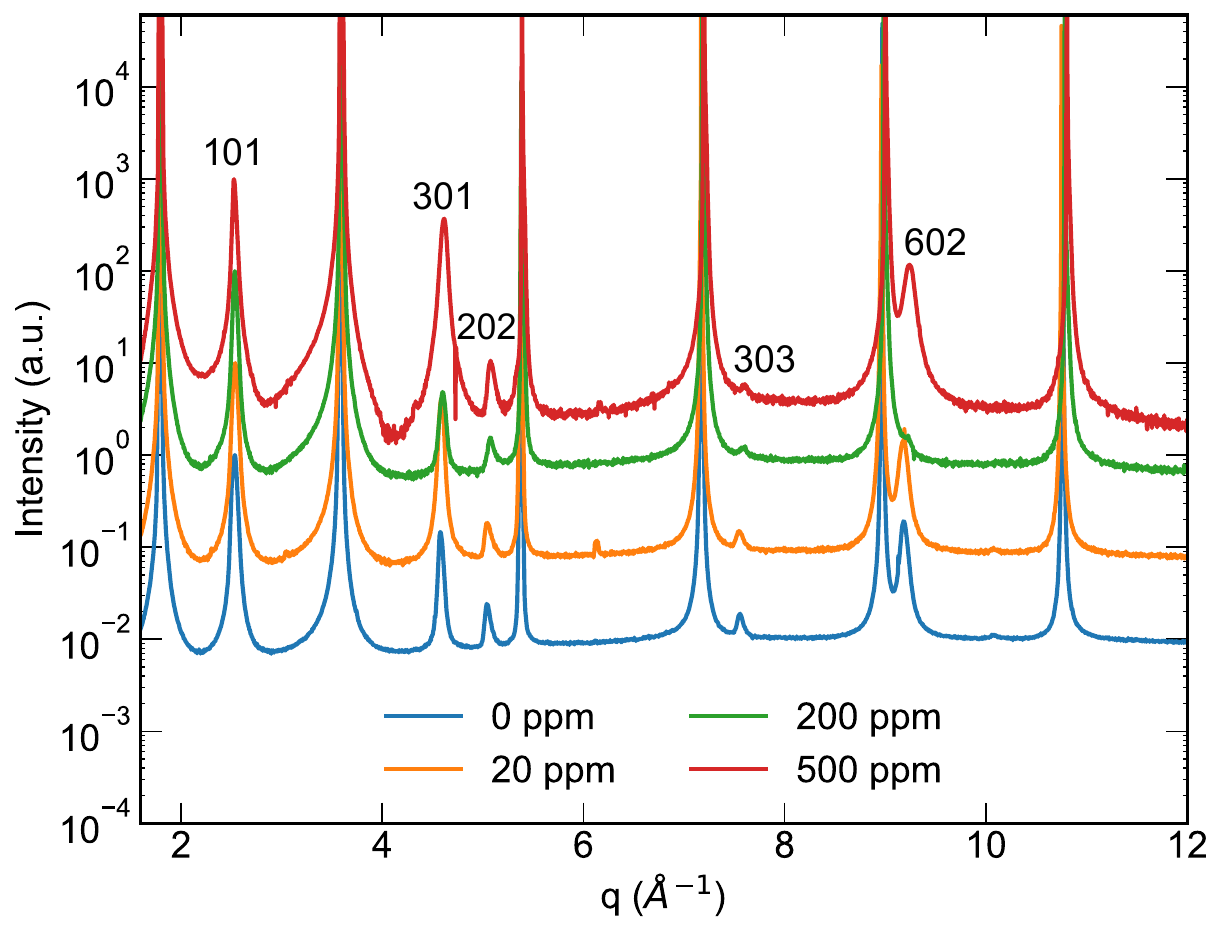}
\caption{(101) Rutile \ch{TiO2} grown on the (012) surface of sapphire, exhibiting primary (101) orientation with a minor ($<10\%$) phase with a (301) orientation.}
\label{fig:XRD_full}
\end{minipage}
\end{figure}

Figure \ref{fig:XRD_full} shows the single crystal X-ray diffraction spectra, confirming the (012)$_{Al_2O_3}$ direction normal to the r-plane surface through the specular (0, K, 2K) scan. The film shows a main peak of (101)$_{TiO_2}$ film orientation on the r-plane surface consistent with previous literature \cite{EngelHerbert:2009p149,Huang:2002p735}. Position of the peaks are roughly equivalent with the expected bulk values indicating that the films are relaxed. The lattice parameter approximates that of bulk \ch{TiO2}. As was noted in previous work\cite{EngelHerbert:2009p149}, the matching of the symmetry between the (012)$_{Al_2O_3}$ and (101)$_{TiO_2}$ planes, leads to a twinned structure with alternating domains on 10's of nm lengthscales. These domains likely lead to a relaxation to the bulk lattice values due to the high-density of domain walls.
Additionally, all films exhibit a minority (301) orientation, which may result from the high density of twin domains. No systematic behavior with doping was observed, suggesting that the minority (301) orientation is unrelated to doping.

\subsection{Impact of erbium concentration on optical properties}
\begin{figure}[ht!]
  \centering
  \begin{subfigure}[b]{0.63\columnwidth}
        \includegraphics[width=\linewidth]{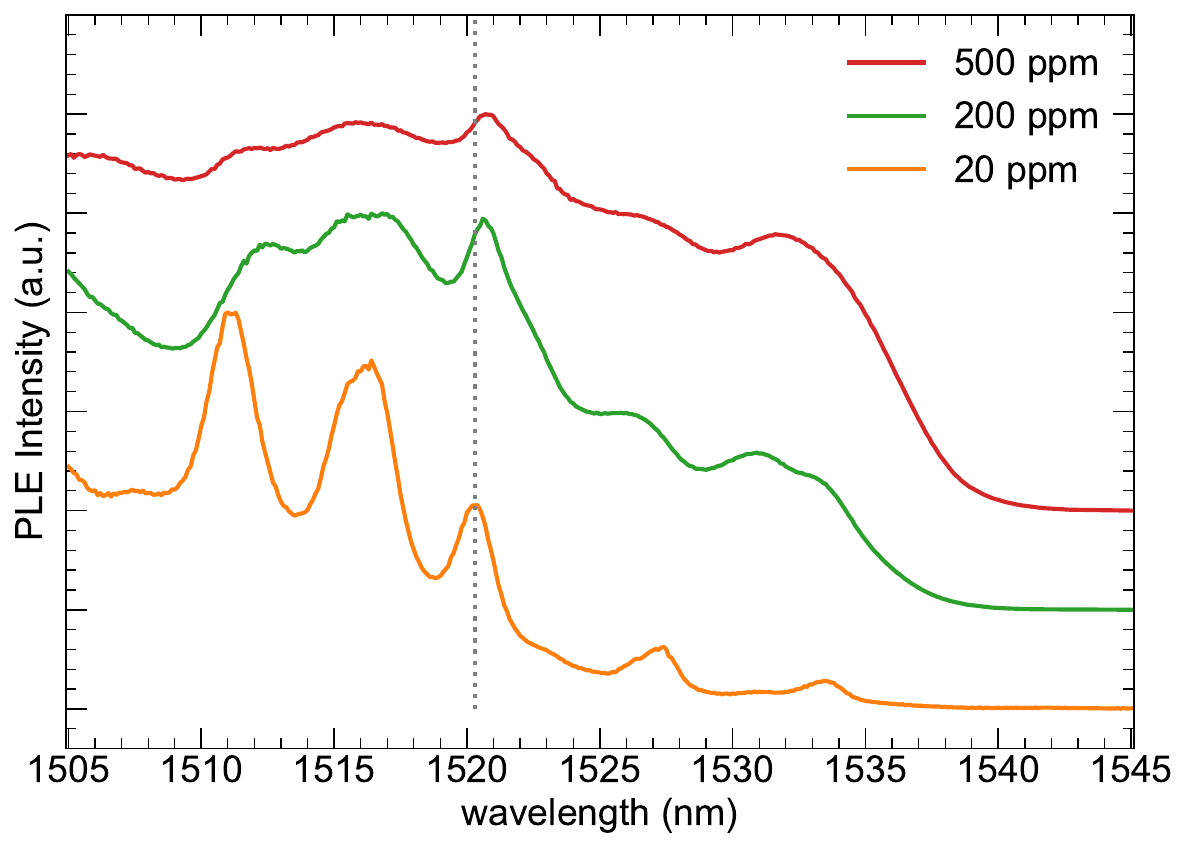}
        \caption{}
        \label{fig:PLE_rutile_doped_series}
    \end{subfigure}%
    \hspace{0pt}  
      \begin{subfigure}[b]{0.36\columnwidth}  
        \includegraphics[width=\linewidth]{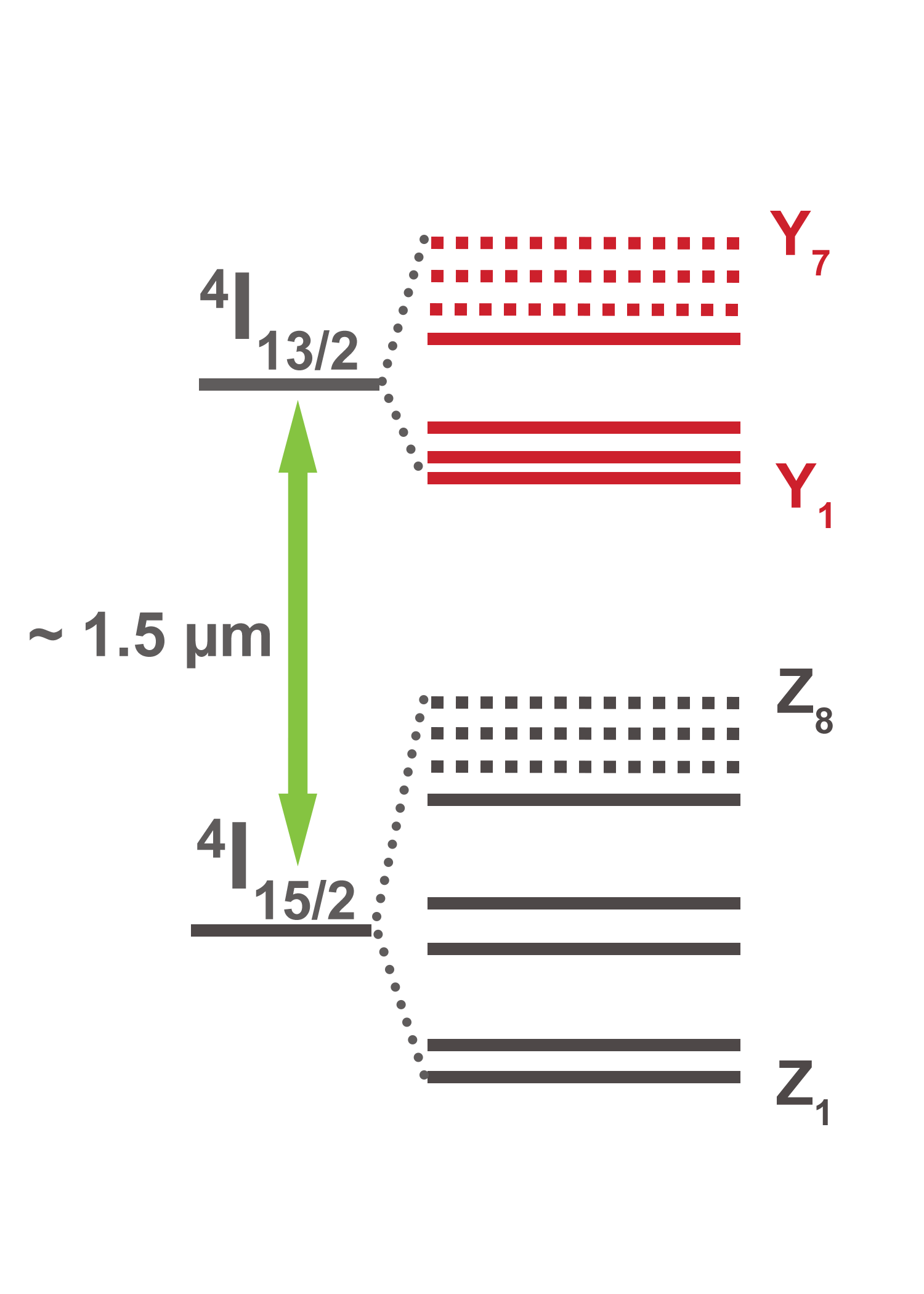}
        \caption{}
        \label{fig:ErbiumTransition}
    \end{subfigure}
\caption{(a) PLE spectra for the set of Er-doped \ch{TiO2} thin films. (b) Energy level diagram for $^4I_{15/2}$ $\rightarrow$ $^4I_{13/2}$ transition in \ch{Er^3+} ion.}
\end{figure}

The impact of increasing \ch{Er^3+} dopant density on the rutile \ch{TiO2} host lattice was investigated by optical characterization of the thin film samples performed via time-resolved PLE spectroscopy. Figure \ref{fig:PLE_rutile_doped_series} shows the resulting absorption spectra near 1.5 µm resonance for the series of single crystal erbium-doped rutile \ch{TiO2} films. The ground $(^4I_{15/2}$) and excited ($^4I_{13/2}$) electronic states of \ch{Er^3+} in rutile split into eight ($Z_{1-8}$) and seven ($Y_{1-7}$) Kramers' doublets, respectively (Figure \ref{fig:ErbiumTransition}) \cite{Phenicie:2019p8928} with \(D_{2h}\) point symmetry for \ch{Er^{3+}} occupying \ch{Ti^{4+}} sites in rutile \ch{TiO2}. The growing background towards the lower wavelengths (i.e., higher energies) stems from off-resonant absorption.

We focus our discussion on the peak highlighted by a vertical dashed line (approximately 1520 nm), which corresponds to the \ch{Er^3+} $Z_{1} \rightarrow Y_{1}$ transition in rutile-phase \ch{TiO2}. \cite{Phenicie:2019p8928} The narrowest observed inhomogeneous linewidth of approximately 50 GHz is found in the sample doped with 20 ppm erbium and increases with higher dopant concentrations. The broadening of the linewidth can be attributed to electric fields generated by charged defects near the \ch{Er^3+} sites, including negatively charged point defects resulting from \ch{Er^{3+}} substituting \ch{Ti^{4+}} sites (i.e., the negative charge upon replacement of \ch{Ti^{4+}} to \ch{Er^{3+}}), and positively charged oxygen vacancy point defects. \cite{STONEHAM:1969p82,Obregon:2013p298, Perez:2019p108873, Singh:2022p,Minervini:1999p339}

The fluorescence lifetimes at the highlighted peak (Table \ref{tab:Fluorescence_Lifetimes}) exhibit a dopant-driven decreasing trend. This shortening of lifetime is consistent with prior observations in \ch{CeO2} with increasing erbium doping, where the decrease is attributed to the introduction of defects. While the exact nature of these defects remains uncertain, they likely introduce non-radiative or quenching pathways that shorten the fluorescence lifetimes \cite{Grant:2024p21121}. If the defects were strictly localized to individual \ch{Er} sites, their interaction would be limited, and neither significant linewidth broadening nor a pronounced reduction in lifetime would be expected. However, the increased linewidth and shorter fluorescence lifetimes at higher dopant concentrations suggest that the defects surrounding the \ch{Er^{3+}} sites are not confined to the immediate vicinity but instead form more extended distributions. These extended defects likely modify the local crystal field around the \ch{Er} ions, thereby influencing the observed optical properties.

The lifetime of the excited state sets an upper limit for the coherence time. Thus, a shorter fluorescence lifetime indicates a more rapid decay of the excited state, which impacts the persistence of quantum superposition and, in consequence, coherence time of the quantum system. Therefore, the observed dopant-driven decrease in the excited state lifetime is critical in assessing the material's coherence times. Understanding the nuances of local electronic and crystal structures is pivotal for comprehending the alterations in the inherent optical properties of Er-doped rutile \ch{TiO2} with varying erbium concentrations.

\begin{table}
\caption{\label{tab:Fluorescence_Lifetimes} \ch{Er^{3+}} fluorescence lifetimes for Er–doped rutile \ch{TiO2} samples.}
\begin{ruledtabular}
\begin{tabular}{cc}
Er Concentration (ppm) & Lifetime (ms) \\
\hline
20 & 2.1 (3)  \\
200 & 0.99 (2) \\
500& 0.63 (1) \\
\end{tabular}
\end{ruledtabular}
\end{table}

\subsection{Influence of erbium on the electronic structure of \ch{TiO2} host lattice}

\begin{figure*}[ht]
\centering
\begin{subfigure}{.45\textwidth}
  \centering
  \includegraphics[width=.9\linewidth]{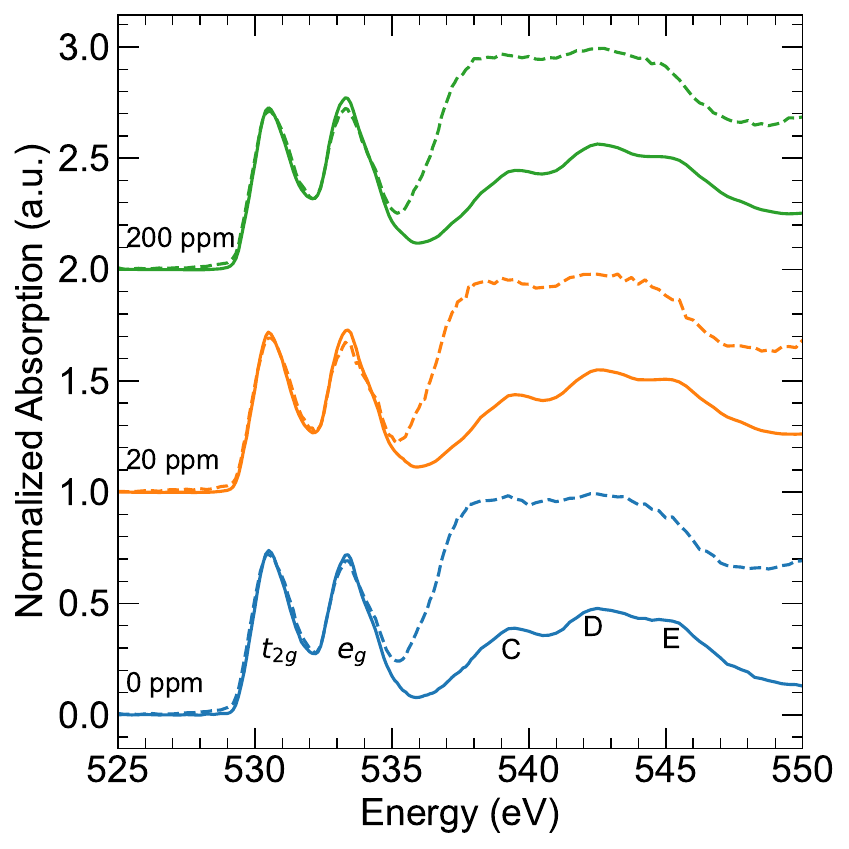}
  \caption{}
  \label{fig:XAS_averaged_OKedge_TEYvsFY}
\end{subfigure}%
\begin{subfigure}{.45\textwidth}
  \centering
  \includegraphics[width=.9\linewidth]{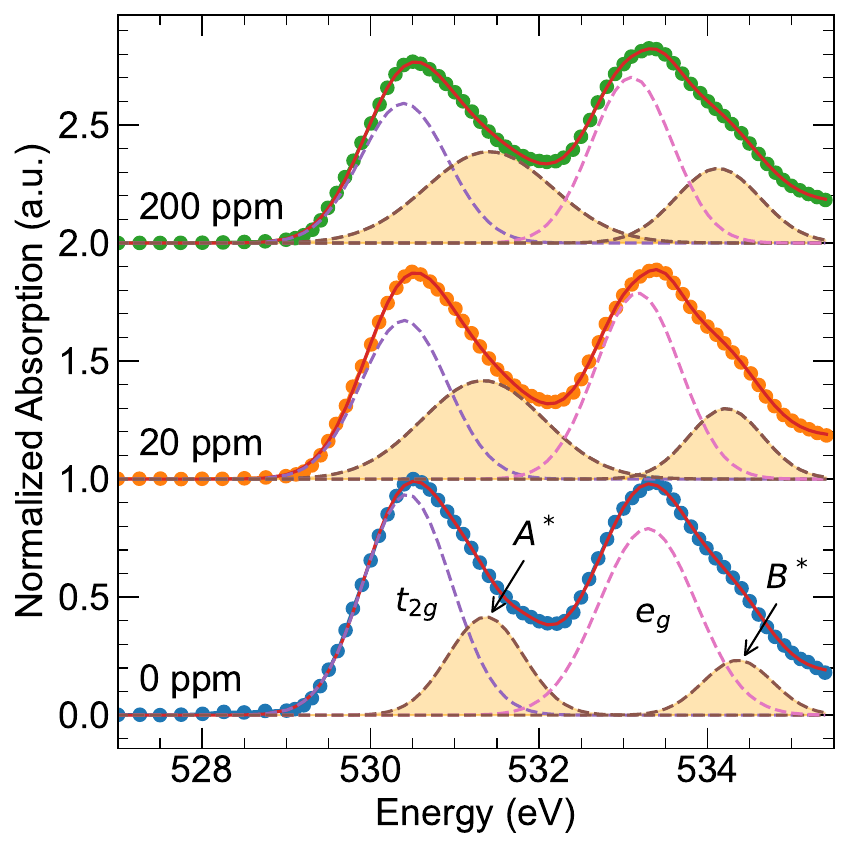}
  \caption{}
  \label{fig:XAS_averaged_OKedge_fits}
\end{subfigure}
\caption{(a) XAS O \textit{K}-edge spectra for undoped (0 ppm) and Er-doped (20 and 200 ppm) \ch{TiO2} films, measured via surface-sensitive TEY mode (solid lines) and bulk-sensitive FY mode (dashed lines). (b) Gaussian fit analysis of TEY spectra depicting pre-peaks \textit{A} and \textit{B}, corresponding to electron transitions from \ch{O} 1s shell into unoccupied \ch{O} 2p orbitals hybridized with Ti $t_{2g}$ and $e_{g}$ orbitals, respectively, with adjacent features $A^*$ and $B^*$ indicating oxygen vacancies.\cite{Vasquez:2016p235209}}
\label{fig:XAS_OKedge}
\end{figure*}

To further elucidate these modifications, we employed XAS to investigate erbium-induced changes in the local electronic structure within the crystal lattice. Starting with oxygen spectroscopy, the O \textit{K}-edge spectra displayed in Figure \ref{fig:XAS_averaged_OKedge_TEYvsFY} are a result of electronic transitions from O(\textit{1s}) states to unoccupied O(\textit{2p}) states. The spectra were acquired in both total electron yield (TEY) and fluorescence yield (FY) modes, represented by solid and dashed lines respectively. The primary distinction between TEY and FY acquisition modes lies in their depth sensitivity, with TEY being more surface-sensitive and FY more bulk-sensitive. In terms of symmetry, \ch{TiO2} consists of a central \ch{Ti^{4+}} surrounded by six \ch{O^{2-}} ions and the octahedral crystal field splits the Ti(\textit{3d}) orbitals into $t_{2g}$ and $e_{g}$ states \cite{Cromer:1955p4708,Ruus:1997p199,Harada:2000p12854}. Geometrically, the $e_{g}$ orbitals from Ti point toward the O \textit{2p} orbitals, while the $t_{2g}$ lobes point between the O atoms leading to strong hybridization between O(\textit{2p}) and Ti(\textit{3d}) orbitals. The significant hybridization leads to two distinguishable pre-peaks at approximately 530 eV and 533 eV labeled as $t_{2g}$ and $e_{g}$ respectively. Moreover, three distinct features at higher energies, with peaks around 539 eV, 542 eV, and 545 eV, denoted as C, D, and E, correspond to delocalized \textit{O}(\textit{2p})–\textit{Ti}(\textit{4s}) and \textit{O}(\textit{2p})–\textit{Ti}(\textit{3d}) hybridized states and are a fingerprint of the rutile phase in \ch{TiO2} \cite{Jiang:2003p341}. 

Two notable trends emerge from comparing the TEY and FY spectra. Firstly, the intensity ratio of the \(t_{2g}\) and \(e_{g}\) pre-peaks varies with increasing erbium concentration in both TEY and FY spectra, suggesting that these variations are not confined to the surface and exist in the bulk of the film as well. Secondly, the FY spectra, which provide greater depth sensitivity than TEY, display broader peaks between 538 and 545 eV. This broadening is attributed to contributions from the sapphire substrate \cite{Hsieh:2014p547}. Due to the strong hybridization between \ch{O}(\textit{2p}) and \ch{Ti}(\textit{3d}) orbitals, the $t_{2g}$ and $e_{g}$ pre-peaks in the \ch{O} \textit{K}-edge spectra are expected to be highly sensitive to changes in the electronic structure. Thus, the fitting analysis illustrated in Figure \ref{fig:XAS_averaged_OKedge_fits} focused on the two pre-peaks in the TEY spectra to gain insights into the electronic structure of the host lattice. Gaussian functions were fitted to the spectra, revealing two distinct features labeled as $A^*$ and $B^*$, adjacent to the main $t_{2g}$ and $e_{g}$ peaks, respectively. These adjacent features are approximately 1 eV blue-shifted from the main peaks and can be attributed to modifications in the electronic structure of the rutile lattice related to oxygen vacancies \cite{Vasquez:2016p235209}. 

Notably, the $A^*$ and $B^*$ features are present even in the undoped sample, indicating non-stoichiometric \ch{TiO2}. Their area increases with erbium doping, as detailed in Table \ref{tab:xasOKedgeFitAnalysis}. Native point defects, such as neutral or charged oxygen vacancies, are common in wide-bandgap metal oxides like \ch{TiO2} \cite{Angelis:2014p9708,Seebauer:2006p57}. The increase in $A^*$ and $B^*$ areas with doping suggests that erbium introduces new oxygen vacancies into the lattice. Each \ch{Er} ion is expected to create only half an oxygen vacancy to balance the charge (i.e., one \ch{O_v^{2+}} vacancy balances the charge of two \ch{Er^{3+}} ions). Despite this, the substantial increase in $A^*$ and $B^*$ peak areas implies that defects around different \ch{Er} sites likely overlap or interact. This overlap leads to significant changes in the XAS spectra, even at low \ch{Er} concentrations, suggesting that the defects are likely not isolated but form a more extensive network within the material. Moreover, the erbium-driven increase in $A^*$ and $B^*$ areas is accompanied by a decrease in the $t_{2g}$ and $e_{g}$ peak areas, reflecting disruptions in the local electronic structure. The creation of new oxygen vacancies disrupts the strong hybridization between \ch{O}(\textit{2p}) and \ch{Ti}(\textit{3d}) orbitals, leading to reduced overlap and thus smaller $t_{2g}$ and $e_{g}$ peak areas.

\begin{table}
\caption{\label{tab:xasOKedgeFitAnalysis} Peak areas obtained from the Gaussian fitting analysis of pre-peaks in XANES \ch{O} \textit{K}-edge spectra. Two main trends are observed. First, the overall decrease in areas of the main $t_{2g}$ and $e_{g}$ pre-peaks with increasing erbium concentration. In contrast, erbium-driven increasing in the $A^*$ and of $B^*$ areas suggest generation of oxygen vacancies within the host lattice.}
\begin{ruledtabular}
\begin{tabular}{ccccc}
Er Concentration (ppm) & \multicolumn{4}{c}{Peak Area (a.u.)} \\
 & $t_{2g}$ & $A^*$ & $e_g$ & $B^*$ \\
\hline
0 & 1.22 (14) & 0.47 (9) & 1.13 (8) & 0.24 (7) \\
20 & 0.86 (14) & 0.76 (9) & 0.98 (8) & 0.33 (7) \\
200& 0.77 (14) & 0.73 (9) & 0.84 (8) & 0.40 (7) \\
\end{tabular}
\end{ruledtabular}
\end{table}



\begin{figure}[ht]
\begin{minipage}[b]{0.95\linewidth}
\centering
\includegraphics[width=\textwidth]{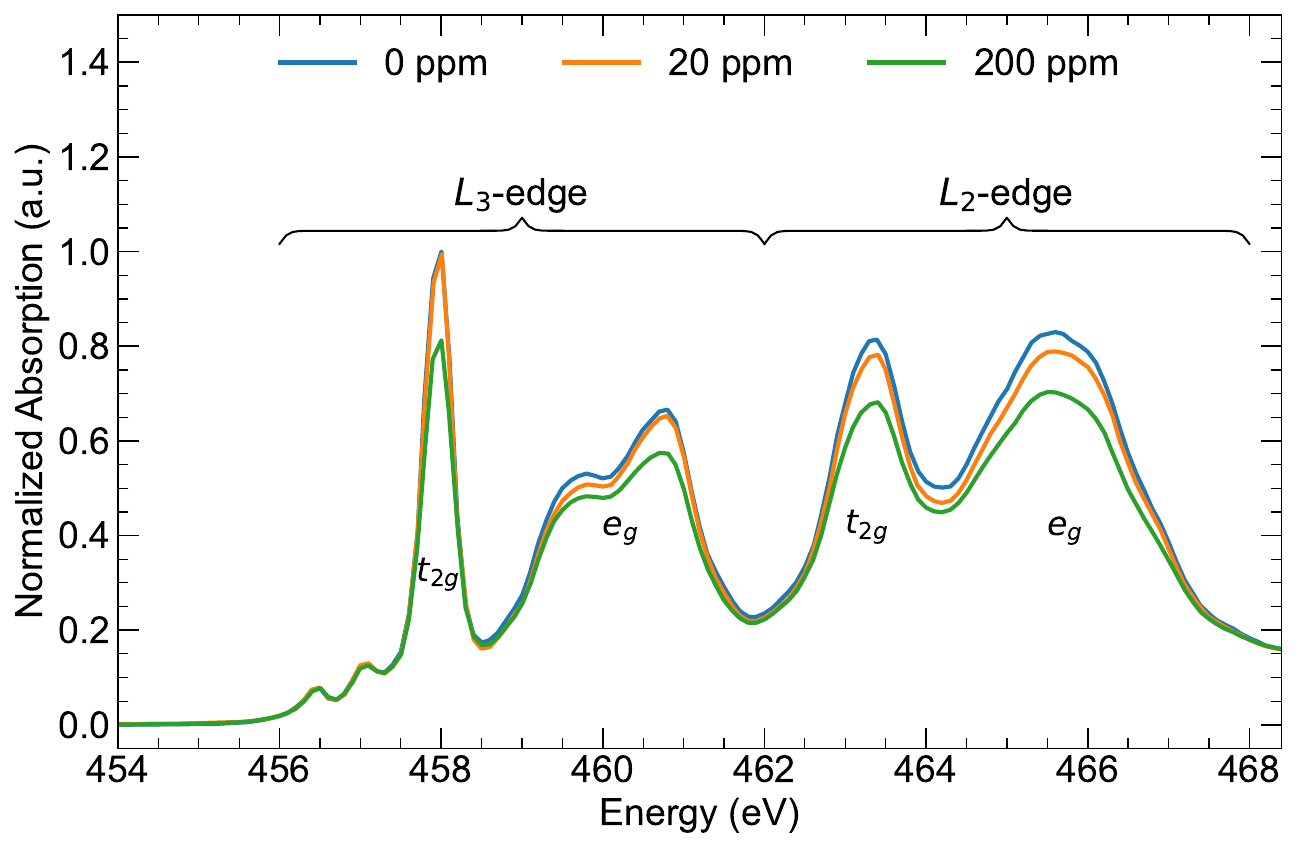}
\caption{Ti L-edge spectra for undoped (0 ppm) and Er–doped \ch{TiO2} samples (20 ppm and 200 ppm).}
\label{fig:xas_TiLedge_normalIncidence}
\end{minipage}
\end{figure}

Similarly to the O \(K\)-edge, the Ti \(L\)–edge XAS spectra are sensitive to changes in electronic structure due to the hybridization of \ch{O}(\textit{2p}) and \ch{Ti}(\textit{3d}) orbitals. Therefore, the Ti \(L\)–edge was probed using TEY mode to investigate possible dopant-induced modifications in the electronic structure (Figure \ref{fig:xas_TiLedge_normalIncidence}). The fine structure in the \(L_3\) and \(L_2\) edges results from the breaking of degenerate electronic orbitals by the lattice crystal field, leading to the splitting of 3d states into three-fold \(t_{2g}\) and two-fold degenerate $e_g$ states. Further distortion from the octahedral lattice induces additional splitting, particularly evident at the \(L_3\)-edge. The first two weak pre-edge peaks are associated with combined effects of particle-hole coupling and crystal field splitting \cite{Groot:1990p928,Krueger:2010p125121}. These features exhibit low intensity due to the incomplete screening of the core hole and the influence of multiplet effects, which reduce the transition intensity. The intense, narrow peak around 458 eV corresponds to the \(t_{2g}\) state, while the split \(e_g\) states produce two peaks associated with the \ch{TiO2} crystal phase.

Expected consequences of doping include charge compensation through the formation of additional defects, such as oxygen vacancies and \ch{Ti^3+}, as well as distortions of the host lattice. \cite{Minervini:1999p339,Li:2005p155315,Vasquez:2018p8740} Prior investigations on \ch{Cr}-doped rutile \ch{TiO2} highlighted the critical role of oxygen point defects in stabilizing \ch{Cr^3+} ions, particularly through the formation of complex defects involving two \ch{Cr} atoms and one oxygen vacancy \cite{Vasquez:2018p8740}. This stabilization is associated with distortions in the crystal lattice due to the elongation of \ch{Ti-O} bonds, which affect the \ch{Ti} $e_g$ orbitals. These orbitals are highly sensitive to changes in the local environment because they interact directly with the \textit{2p} orbitals of the surrounding \ch{O} atoms. As the local symmetry and crystal field around the \ch{Ti} ions change, the energy levels of the $e_g$ orbitals shift accordingly. In our study, the low concentration of \ch{Er^3+} dopant—on the order of hundreds of parts per million (ppm)—makes it challenging to observe significant lattice distortions or prominent features associated with $e_g$ orbital splitting or \ch{Ti^{3+}}-related signals. However, comparisons between undoped and doped samples reveal an overall decrease in spectral intensity with \ch{Er} doping, suggesting that the dopant induces changes in the local electronic environment around the \ch{Ti} ions. These changes may arise from slight distortions in the \ch{Ti-O} coordination environment, which in turn alter the crystal field and electronic structure of the \ch{Ti} sites.

\begin{figure}[ht]
\begin{minipage}[b]{0.95\linewidth}
\centering
\includegraphics[width=\textwidth]{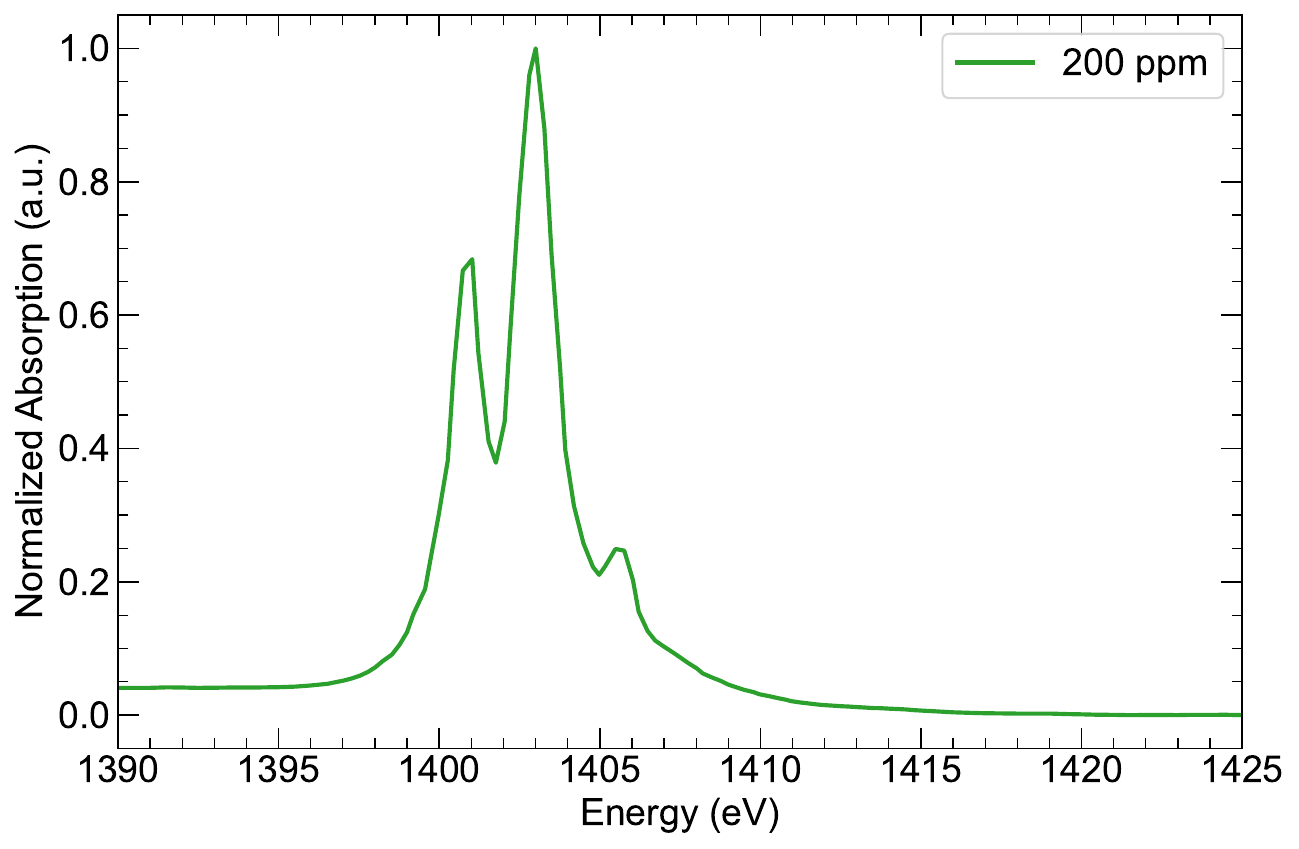}
\caption{XAS spectrum at the Er $M_5$ edge for 200 ppm doped Er:\ch{TiO2}.}
\label{fig:ErM5_XAS}
\end{minipage}
\end{figure}

To gain deeper insight into how the dopant affects the formation of additional defects within the host lattice Erbium spectroscopy was performed. The XAS spectrum at the Er \(M_5\)-edge (Figure \ref{fig:ErM5_XAS}) exhibits three distinct peaks corresponding to \(3d \rightarrow 4f\) transitions. These transitions, characterized by changes in the total angular momentum quantum number (\(\Delta J = 0, \pm 1\)), are indicative of erbium in the trivalent state \cite{Thole:1985p5107}. 

\subsection{Structural distortions upon erbium incorporation into \ch{TiO2}}

\begin{figure*}[htbp]  
    \centering
    \begin{minipage}[b]{\textwidth} 
        \centering
        \begin{subfigure}{0.32\linewidth}  
            \includegraphics[width=\linewidth]{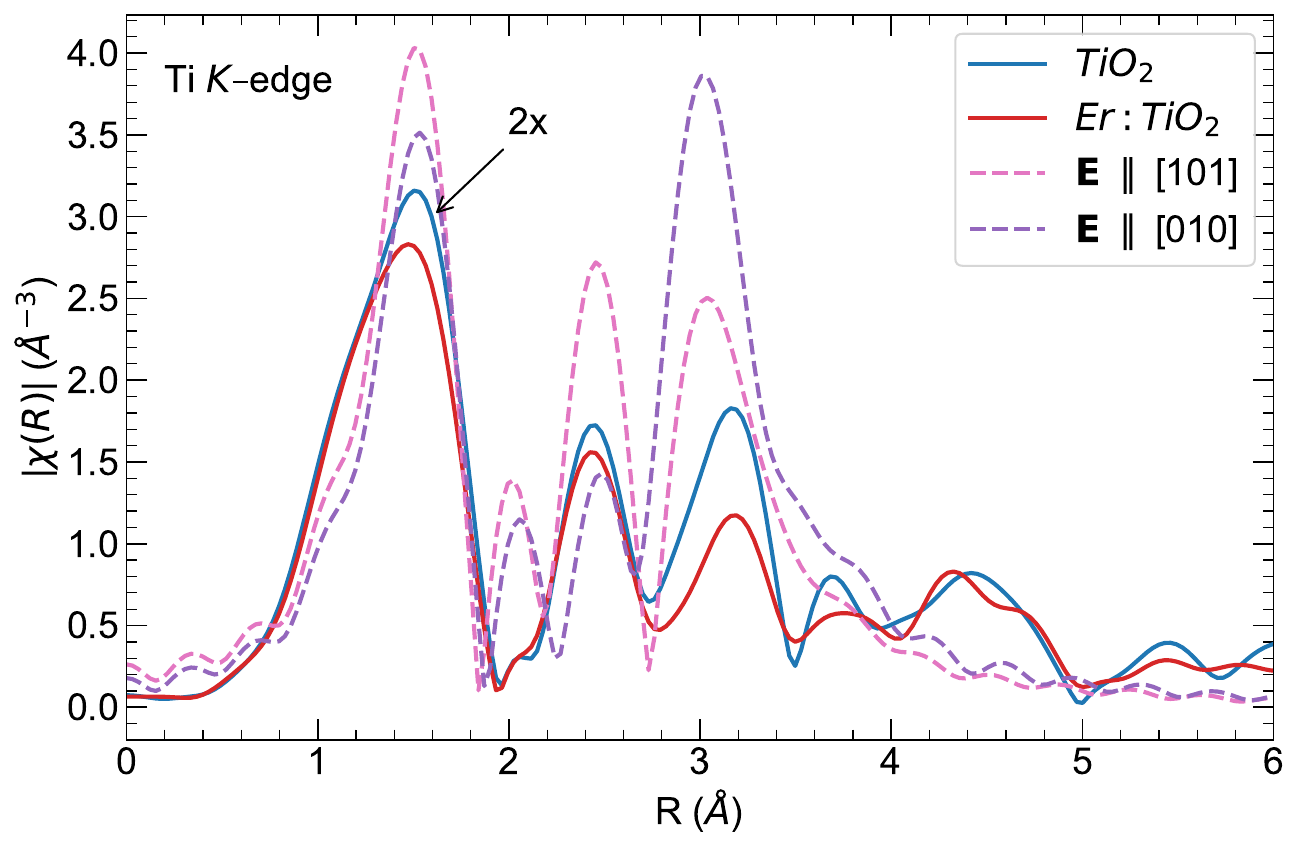} 
            \caption{}
            \label{fig:TiPolarizationsSimulation}
        \end{subfigure}\hfill  
        \begin{subfigure}{0.32\linewidth}
            \includegraphics[width=\linewidth]{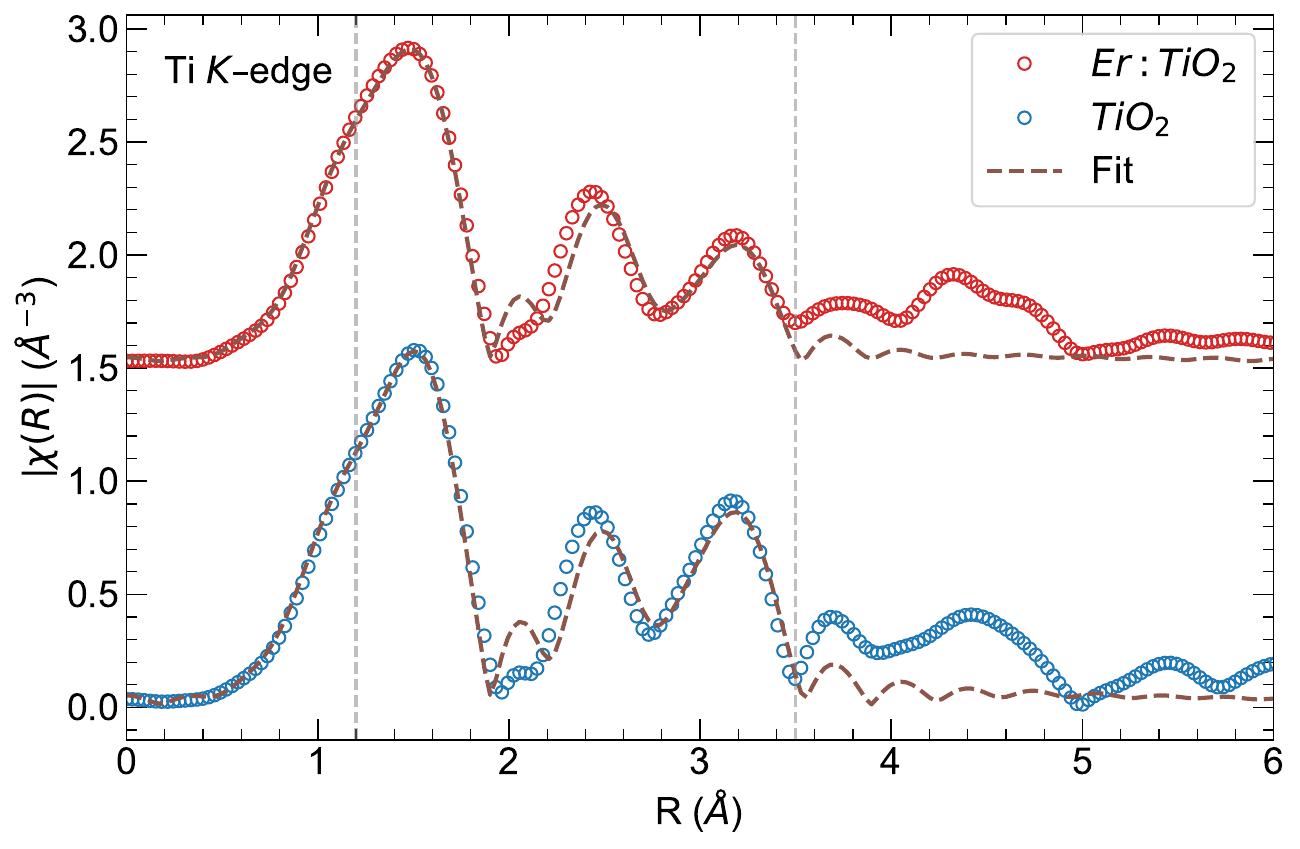} 
            \caption{}
            \label{fig:FT_Ti_TiO2vsSample}
        \end{subfigure}\hfill
        \begin{subfigure}{0.32\linewidth}
            \includegraphics[width=\linewidth]{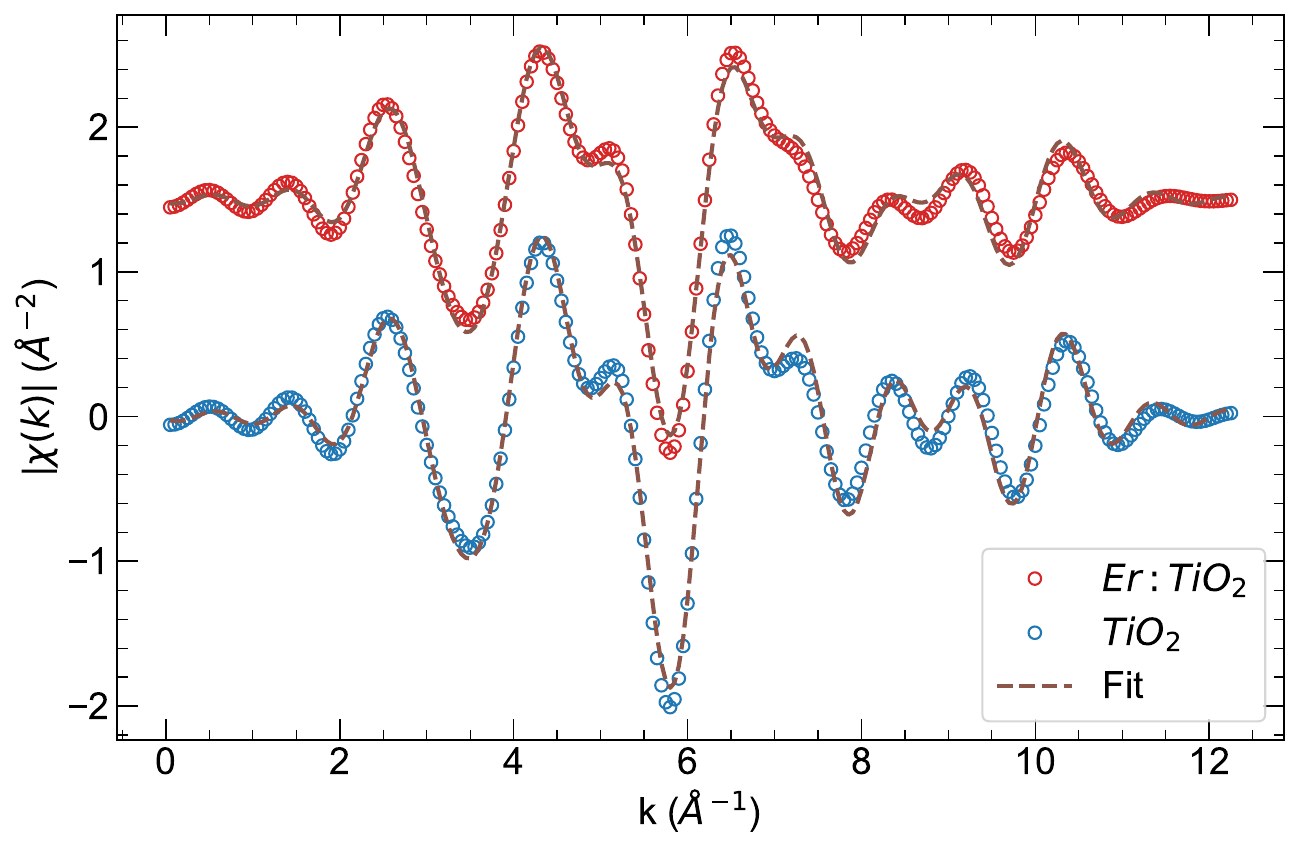} 
            \caption{}
            \label{fig:Chi_k_Ti_TiO2vsSample}
        \end{subfigure}
        
        \vspace{3mm} 
        \begin{subfigure}{0.32\linewidth}
            \includegraphics[width=\linewidth]{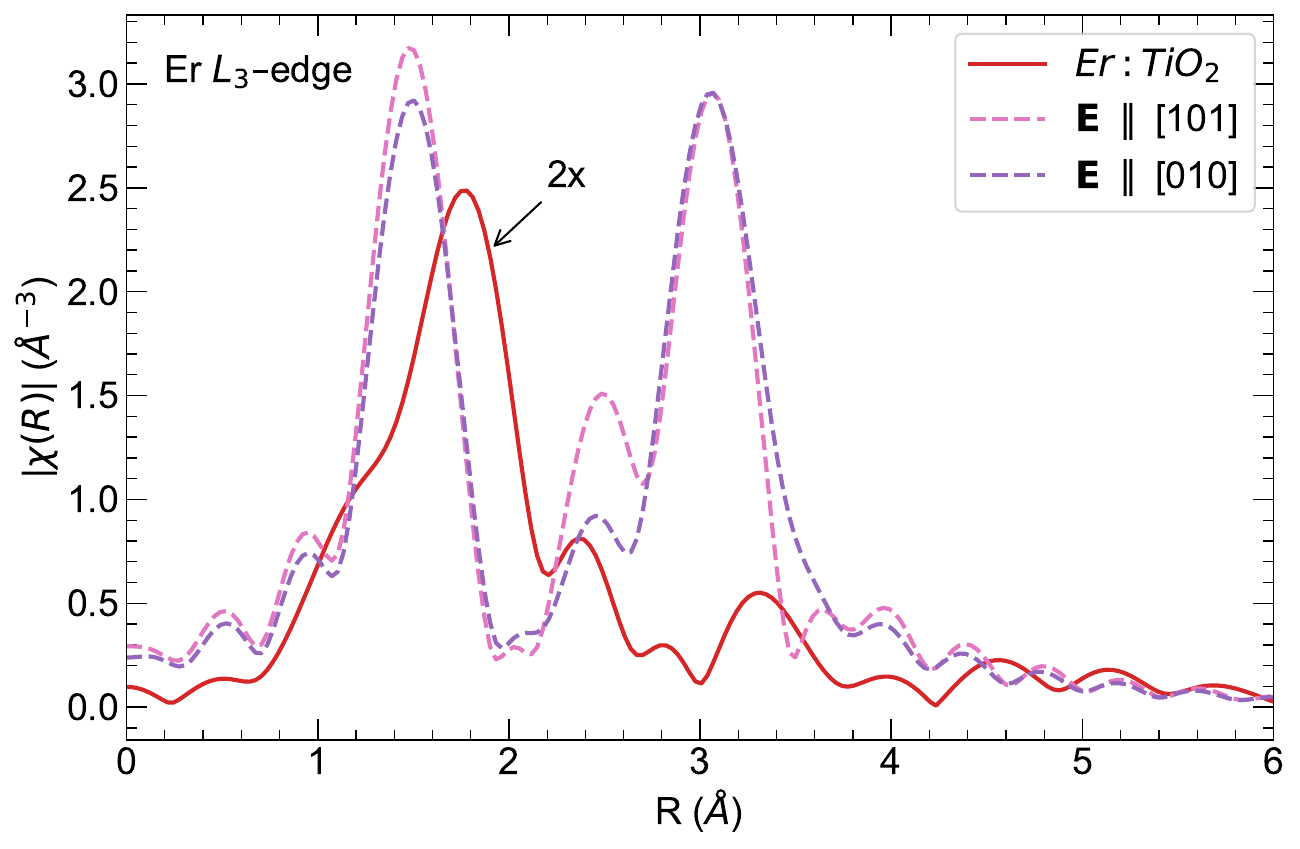} 
            \caption{}
            \label{fig:ErPolarizationsSimulation}
        \end{subfigure}\hfill
        \begin{subfigure}{0.32\linewidth}
            \includegraphics[width=\linewidth]{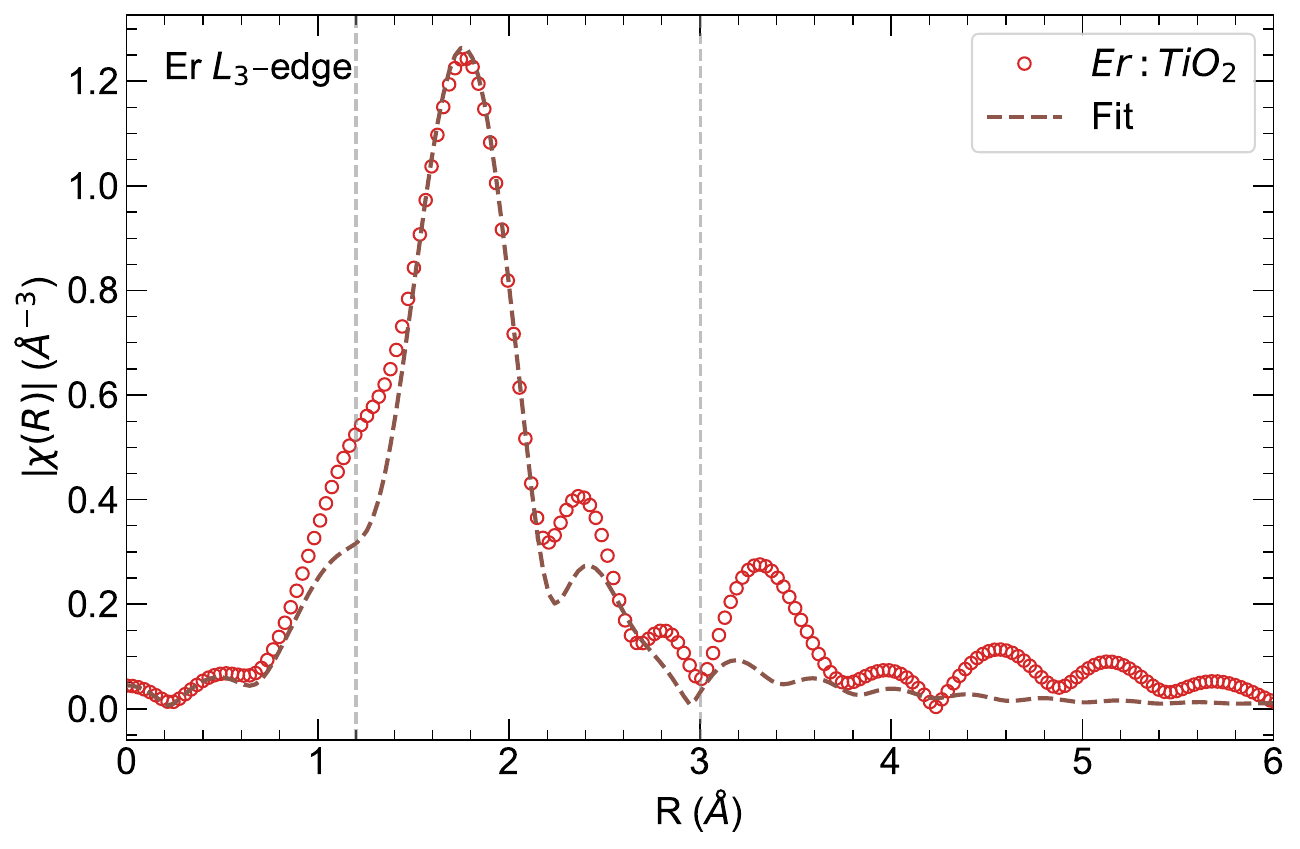} 
            \caption{}
            \label{fig:FT_Er_TiO2vsSample}
        \end{subfigure}\hfill
        \begin{subfigure}{0.32\linewidth}
            \includegraphics[width=\linewidth]{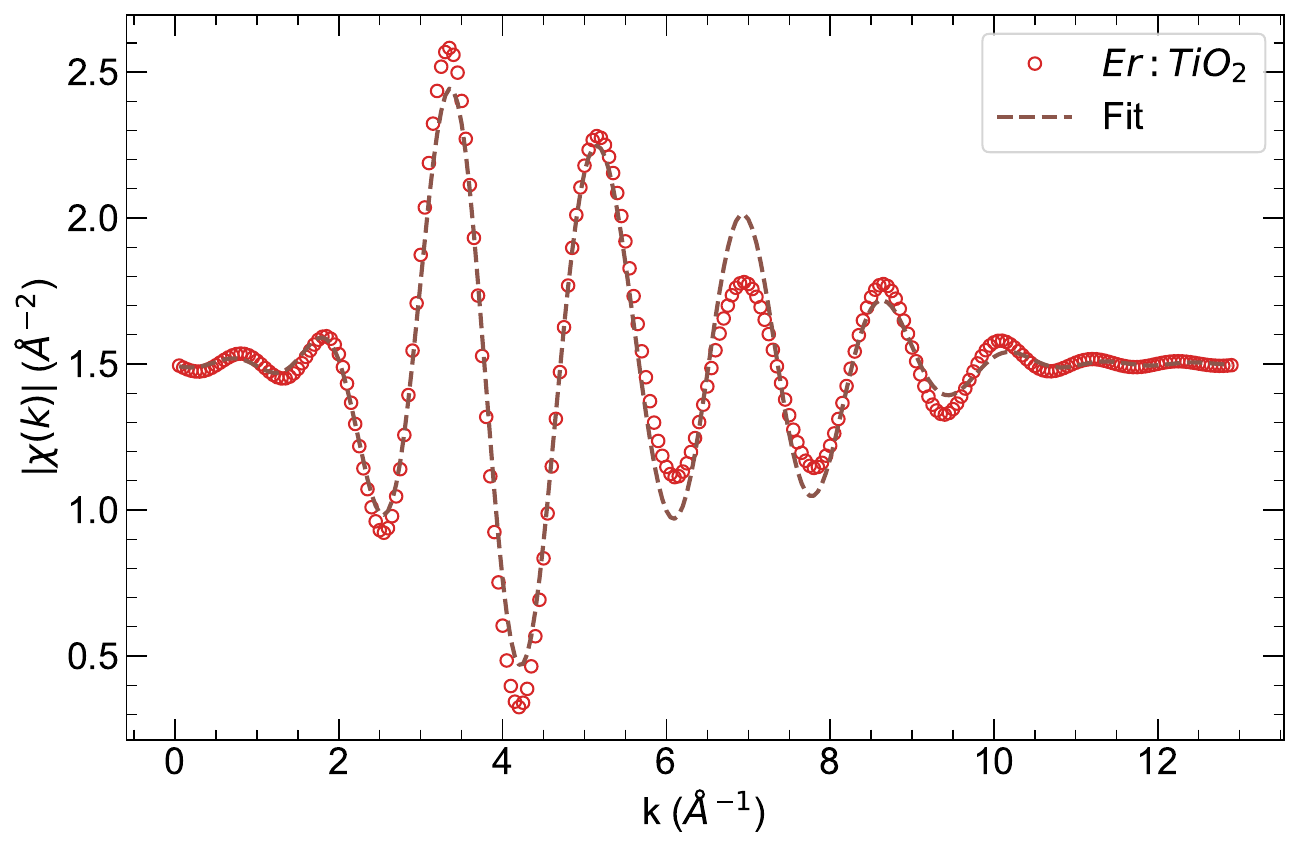} 
            \caption{}
            \label{fig:Chi_k_Er_TiO2vsSample}
        \end{subfigure}
    \end{minipage}%
    \caption{Fourier-transformed (a) Ti \textit{K}-edge and (d) Er \(L_3\)-edge XAS spectra of undoped (solid blue) and 500 ppm Er-doped (solid red) \ch{TiO2} samples. The $k$ range used in the FTs was [2-11]~\AA$^{-1}$ and [2-10]~\AA$^{-1}$ for Ti $K$ and Er $L_3$-edges, respectively. The dashed curves represent FEFF simulations for electric field polarizations along the [010] and [-101] directions. Fitted Fourier-transformed (b) Ti \textit{K}-edge and (c) Er \(L_3\)-edge spectra for undoped (blue circles) and 500 ppm Er-doped \ch{TiO2} (red circles) with vertical dashed lines showing the regions used in the real-space fits. Panels (c) and (f) show the back Fourier transform of data and fits in the real-space region marked by the vertical lines for the Ti \textit{K}-edge and Er \(L_3\)-edge, respectively}
    \label{fig:XAFS_Analysis}
\end{figure*}

The local environment around the \ch{Er^{3+}} sites was further examined using EXAFS at the \ch{Er} \(L\)-edge and \ch{Ti} \(K\)-edge as depicted in Figure \ref{fig:XAFS_Analysis}. A comparison of the experimental data with FEFF simulations is shown in Figures \ref{fig:TiPolarizationsSimulation} and \ref{fig:ErPolarizationsSimulation}, where the Fourier-transformed (FT) XAFS data for the Ti \textit{K}-edge and Er \(L_3\)-edge, respectively, are represented by solid lines.

These comparisons revealed significant differences, particularly in the FT Er \(L_3\)-edge spectra, where a marked shift towards higher distances in the first shell was observed, along with a notable suppression in the higher-R structure. These observations suggest alterations in the local structure around erbium, such as bond length variations or changes in coordination numbers at greater distances. The experimental data, acquired at 300 K, were scaled by a factor of 2 in these two plots to better match with simulations conducted at 0 K.

Figures \ref{fig:FT_Er_TiO2vsSample} and \ref{fig:FT_Ti_TiO2vsSample} display FT XAFS data and fits for Er \(L_3\)-edge and Ti \textit{K}-edge, respectively. Each polarization contribution was equally weighted (0.5) in the fits to account for in-plane polarization and sample spin, employed during the data acquisition to reduce Bragg peak artifacts in the experiment. The analysis involved simultaneous fitting of data weighted by different $k$-weights to minimize correlations between coordination number and the Debye-Waller factor ($\sigma^2$). The small nominal split of 0.04 \AA~ in first shell Ti/Er-O distances cannot be resolved with the limited $k$-range of the XAFS data ($k_{\rm max}\sim 10 $ \AA$^{-1}$) hence the fitted bond length correction was forced to be proportional to the nominal bond lengths. Similarly, coordination numbers of split distances were fitted while maintaining their ratio fixed to nominal values. The FT Er \(L_3\)-edge spectrum (Figure \ref{fig:FT_Er_TiO2vsSample}) revealed erbium coordinated to oxygen for the first shell (\ch{Er-O}), with an expanded bond length of 2.20 Å (Table \ref{tab:FTFitAnalysis}). This distance is significantly larger than the corresponding \ch{Ti-O_{(1)}} bond length of 1.92 Å observed in the \ch{Ti} \textit{K}-edge analysis, indicating a local expansion in the lattice around the erbium dopant. Additionally, the presence of \ch{Ti} neighbors at a distance of 3.09 Å in the second shell suggests that \ch{Er^3+} is occupying \ch{Ti^4+} lattice sites. 

The significant local expansion observed around \ch{Er} aligns with the difference in ionic radii between \ch{Er^3+} and \ch{Ti^4+}. The theoretical difference in ionic radii is 0.29 Å (\(\mathrm{0.89 \, \text{\AA} - 0.605 \, \text{\AA}}\) \cite{Shannon:1969p925}), which closely matches the experimental local expansion of 0.28 Å, as indicated by the difference between the \ch{Er-O_{(1)}} bond length (2.20 Å) and the \ch{Ti-O_{(1)}} bond length (1.92 Å) listed in Table \ref{tab:FTFitAnalysis}. This close agreement between theoretical and experimental values, combined with nominal (N=2) \ch{Er-Ti} coordination in the second shell at a bond distance of 3.09 Å, supports the hypothesis that \ch{Er^3+} is substituting at \ch{Ti^4+} sites, albeit with significant lattice distortion especially at higher order shells. Interestingly, the amplitude of Ti $K$-edge XAFS is somewhat reduced in the doped sample, despite the rather low doping level of 500 ppm, indicating that structural distortions around Er ions are extended, affecting a larger \% of Ti ions. These results can also be correlated to the overall intensity decreasing in \ch{Ti} \(L\)-edge XAS spectra with doping portrayed in Figure \ref{fig:xas_TiLedge_normalIncidence}, as erbium ions appear to cause lattice distortions. The Ti \textit{K}-edge analysis (Figure \ref{fig:FT_Ti_TiO2vsSample}) for the undoped sample (blue circles) yielded fitted distances closely resembling the nominal \ch{TiO2} structure. Small differences were observed for the doped sample (red circles), although signatures of increased disorder are present particularly at the longer Ti-Ti correlations (Table \ref{tab:FTFitAnalysis}). The same amplitude reduction factor (S\(_0^2\)) of 0.93 was used for both the Ti \textit{K}-edge and Er \(L_3\)-edge fits to ensure consistency across the analyses. 

The substitution of \ch{Ti^4+} by \ch{Er^3+} ions raises important questions about the mechanisms of charge compensation that stabilize the dopant within the host structure. The increased area of the \(A^*\) and \(B^*\) features in the \ch{O} \(K\)-edge spectra, which are associated with oxygen vacancies, suggests that these vacancies play a key role in compensating for the charge imbalance introduced by \ch{Er^3+} ions, thereby maintaining charge neutrality \cite{Luo:2011p3046, Rao:2019p12321}. EXAFS analysis indicates that the first coordination shell of both \ch{Er^3+} and \ch{Ti^4+} remains fully occupied by oxygen, as reflected by coordination numbers close to 6 for both species. Specifically, the coordination number of \ch{Er-O} is 6.4 (3) (Table \ref{tab:FTFitAnalysis}), while the coordination number for \ch{Ti-O} is 6.2 (3), supporting full occupation in the first shell. However, the error bars of around 5$\%$ in coordination number allow for the presence of oxygen vacancies in the first \ch{Ti-O} shell. Thus, while the first shell appears fully occupied within experimental uncertainty, the possibility of oxygen vacancies beyond the first \ch{Er-O} shell in the lattice cannot be ruled out.

The $\sigma^2$ factors indicate increased disorder beyond the first shell, particularly around the \ch{Er^3+} ions, with a value of 0.0140 (5) \AA$^2$ for the \ch{Er-Ti_{(1)}} distance. Additionally, major differences between the data and simulations in the higher shells (Fig.~\ref{fig:ErPolarizationsSimulation} point to erbium-driven structural changes beyond the first coordination shell, which may include oxygen vacancies located farther from the \ch{Er^3+} sites. We were unable to model the Er local structure beyond the \ch{Er-Ti_{(1)}} distance. The long-range distortions in the local environment around \ch{Er^3+} ions, extending to at least 4 Å, suggest that these defects are not isolated but cause widespread changes in the lattice. Even at 500 ppm doping levels these extensive distortions may overlap affecting a significant amount of Ti atoms, as suggested by the Ti $K$- and $L$-edge probes.

Finally, revisiting the PLE discussion, the observed decrease in fluorescence lifetime of \ch{Er^3+} can be attributed to the dopant-induced generation of oxygen vacancies. These point defects act as non-radiative recombination centers, enabling energy transfer between \ch{Er^3+} ions and oxygen vacancy defects, which leads to a quenching effect. Consequently, the fluorescence lifetime decreases due to enhanced non-radiative recombination pathways. Since the fluorescence lifetime imposes an upper limit on coherence time in quantum systems, the observed reduction in fluorescence lifetimes indicates a corresponding decrease in coherence times. 

\begin{table*}
\caption{\label{tab:FTFitAnalysis}Structural parameters from \ch{Er} and \ch{Ti} XAFS fitting analysis. The amplitude reduction factor (S\(_0^2\)) of 0.93 was determined from \ch{Ti} K-edge fits and fixed in \ch{Er} \(L_3\)-edge fits. FEFF simulations used nominal \ch{TiO2} bond lengths. The footnote values correspond to the nominal \ch{Ti-O} and \ch{Ti-Ti} bond lengths. Lattice distortions were accounted for during fitting, reflecting observed disorder in higher-R shells. Coordination numbers for higher-R shells (i.e., \ch{Ti-Ti_{(2)}} and \ch{Er-Ti_{(1)}}) are absent due to significant disorder and weak scattering amplitudes.}

\begin{ruledtabular}
\begin{tabular}{cccccccc}

Probed Atom & Bond & \multicolumn{2}{c}{Bond Length (\AA)} & \multicolumn{2}{c}{Debye Waller Factor (\AA$^2$)} & \multicolumn{2}{c}{Coordination Number} \\

&  & 0 ppm & 500 ppm & 0 ppm & 500 ppm & 0 ppm & 500 ppm \\ \hline

\multirow{4}{*}{Ti \textit{K}-edge} & \ch{Ti-O_{(1)}} & 1.934 (5)\footnote{1.944 \AA \label{TiO1NominalBondLength}} & 1.92 (1) & 0.0055 (5) & 0.0060 (1) & 4.1 (2) & 4.1 (2) \\

& \ch{Ti-O_{(2)}} & 1.978 (5)\footnote{1.988 \AA \label{TiO2NominalBondLength}} & 1.96 (1) & 0.0055 (5) & 0.0060 (1) & 2.1 (1) & 2.1 (1) \\

& \ch{Ti-Ti_{(1)}} & 3.02 (1)\footnote{2.959 \AA \label{FirstTiTiNominalBondLength}} & 3.00 (3) & 0.0040 (1) & 0.0030 (1) & 2\footnote{Fixed \label{FixedCoordination}} & 2\footref{FixedCoordination} \\
 
& \ch{Ti-Ti_{(2)}} & 3.61 (1)\footnote{3.57 \AA \label{SecondTiTiNominalBondLength}} & 3.62 (2) & 0.0062 (9) & 0.0010 (4) & 8\footref{FixedCoordination} & 8\footref{FixedCoordination} \\
  
\multirow{3}{*}{Er $L_3$-edge} & \ch{Er-O_{(1)}} & - & 2.20 (1) & - & 0.0056 (7) & - & 4.3 (2) \\

& \ch{Er-O_{(2)}} & - & 2.25 (1) & - & 0.0056 (7) & - & 2.1 (1) \\
& \ch{Er-Ti_{(1)}} & - & 3.09 (3) & - & 0.0140 (5) & - & 2\footref{FixedCoordination} \\

\end{tabular}
\end{ruledtabular}
\end{table*}


In conclusion, our investigation into epitaxial \ch{Er}-doped rutile \ch{TiO2} thin films reveals significant effects of erbium doping on both structural and electronic properties. XAS analysis at the \ch{O} \(K\)-edge indicates that erbium doping leads to the formation of oxygen vacancies, which alter the electronic structure of the host lattice. EXAFS analysis for the Er \(L_3\)-edge spectra indicates that \ch{Er^3+} ions substitute for \ch{Ti^4+} ions in the \ch{TiO2} lattice. This substitution results in notable lattice distortions due to the larger ionic radius of \ch{Er^3+} compared to \ch{Ti^4+}. Coordination number analysis reveals consistent six-fold coordination around erbium, suggesting that neighboring oxygen ions stabilize the local charge environment. However, while our findings indicate significant distortions in higher-order shells, which are likely associated with oxygen vacancies displaced from the first shell, further comparative studies are necessary to definitively determine the exact substitution sites. The broadening of the \ch{f-f} excitation observed in PLE suggests a distribution of crystal fields, indicating structural disorder extending beyond the first coordination shell. This disorder impacts the crystal field sensed by the 4f electrons. The lattice distortions and additional point defects are expected to enhance non-radiative energy transfer, which correlates with the reduced fluorescence lifetimes observed by PLE. This underscores the complex interplay between dopant-induced structural modifications and optical properties, emphasizing the importance of detailed local electronic and crystal structure investigations for advancing quantum information processing technologies.

\begin{acknowledgments}
This work was supported by Q-NEXT, a U.S. Department of Energy Office of Science National Quantum Information Science Research Centers under Award No. DE-FOA-0002253. The use of the Advanced Photon Source, Argonne National Laboratory was supported by the U.S. Department of Energy, Office of Science, Basic Energy Sciences, under Contract No. DE-AC02-06CH11357.
\end{acknowledgments}

\section*{Data Availability Statement}

The data that support the findings of this study are available within the article. 

\appendix


\nocite{*}
\bibliography{aipsamp}

\end{document}